\documentclass[twocolumn,prd,amssymb,amsmath,preprintnumbers,floatfix,aps]{revtex4-1}
\usepackage{dcolumn,graphicx,bm,psfrag }

\newcommand{\be}{\begin{equation}}
\newcommand{\ee}{\end{equation}}
\newcommand{\bear}{\begin{eqnarray}}
\newcommand{\eear}{\end{eqnarray}}
\newcommand{\ba}{\begin{array}}
\newcommand{\ea}{\end{array}}
\newcommand{\met}{ E_T\!\!\! \!\! \! \! \slash \;\, } 

\begin{document}
\title{\boldmath \bf \Large A $W'$ boson near 2 TeV: predictions for Run 2 of the LHC } 
\author{\bf Bogdan A. Dobrescu$^\star$ and Zhen Liu$^{\star \diamond}$} 

\affiliation{$\star$ Theoretical Physics Department, Fermilab, Batavia, IL 60510, USA \\ 
$\diamond$ PITT PACC,  Department of Physics and Astronomy, University of Pittsburgh, PA 15260, USA}

\date{\normalsize  June 21, 2015; revised October 5, 2015}

\begin{abstract}
We present a renormalizable theory that includes a $W'$ boson of mass in the 1.8--2 TeV range,
which may explain the excess events reported by the ATLAS Collaboration in a $WZ$ final state, and by the 
CMS Collaboration in  $e^+\!e^- jj$, $Wh^0$ and $jj$ final states. The $W'$ boson couples to right-handed quarks and leptons,
including Dirac neutrinos with TeV-scale masses. 
This theory  predicts a $Z'$ boson of mass in the 3.4--4.5 TeV range. 
The cross section times branching fractions for the narrow
$Z'$ dijet and dilepton peaks at the 13 TeV LHC are 
 10 fb and 0.6 fb, respectively, for $M_{Z'}= 3.4$ TeV,  and an order of magnitude smaller for $M_{Z'}= 4.5$ TeV.
\end{abstract}
\vspace{.4cm}

\maketitle

{\it\bf Introduction.}---The LHC, the highest energy collider built so far, has directly probed the laws of physics at distance
scales as small as  $\sim 5\times 10^{-20}$ m, and over the next few years will extend the exploration 
by another factor of two.
The Standard Model (SM) of particle physics has been spectacularly confirmed through various
analyses based on data obtained during Run 1 of the LHC. 

Recently, though, a few deviations from the SM
predictions have been reported by the ATLAS and CMS Collaborations 
in invariant mass distributions near 2 TeV:   
 \\
{\it 1)} a $3.4 \sigma$ excess at $\sim$2 TeV in the ATLAS  search \cite{Aad:2015owa} 
for a $W^\prime$ boson decaying into $WZ \to JJ$, where $J$ stands for a wide jet formed by the 
two nearly colinear jets produced in the decays of a boosted $W$ or $Z$ boson. 
The mass range with significance above $2\sigma$ is $\sim 1.9$--2.1 TeV; the global significance is $2.5\sigma$.
A CMS search \cite{Khachatryan:2014hpa} for $JJ$ resonances,
without distinguishing between the $W$- and $Z$-tagged jets, has a $1.4 \sigma$ excess at $\sim$1.9 TeV.
 \\ 
{\it 2)} a $2.8\sigma$ excess in the 1.8 -- 2.2 TeV bin in the CMS search~\cite{Khachatryan:2014dka} for a $W'$  
and a heavy ``right-handed" neutrino, $N_R$, through the $W^\prime \to N_R \,e \to eejj$ process. \\  
{\it 3)} a $2.2\sigma$ excess in the 1.8 -- 1.9 TeV bin
in the CMS search \cite{CMS:2015gla} for $W^\prime \to Wh^0$, where the SM Higgs boson, $h^0$, is highly boosted and 
decays into $b\bar b$, while $W \to \ell \nu$.   \\  
{\it 4)} a $\sim 2\sigma$ excess at $\sim$$1.8$ TeV in the CMS dijet resonance search \cite{Khachatryan:2015sja}. 
The ATLAS search \cite{Aad:2014aqa} in the same channel has yielded only a $1\sigma$ excess at $1.8$ TeV.

Although none of these deviations is significant enough to indicate a new phenomenon, it behooves us to 
inquire whether  a self-consistent theory may explain all of them.
Here we construct a renormalizable theory that explains quantitatively these deviations, and derive its predictions for signals that can be probed in Run 2
of the LHC. 

The deviations showed up in searches for a $W'$ boson but several theoretical and experimental hurdles 
need to be overcome before a particle of mass near 2 TeV can be inferred.
The $eejj$ excess suggests that the  $W'$ boson couples to right-handed fermions, as in left-right symmetric models \cite{Mohapatra:1986uf}.
However, those models predict a Majorana mass for $N_R$, so the number of events with same-sign lepton pairs should be approximately equal
to that for opposite-sign lepton pairs \cite{Keung:1983uu} (except for the case where two $N_R$'s with 
CP violating mixing are degenerate \cite{Gluza:2015goa}). 
As the CMS excess consists almost entirely of $e^+e^-$ pairs, we will extend the left-right symmetric models in order to allow a TeV-scale Dirac mass for $N_R$.
 
Another issue is  that all gauge extensions of the SM that include a $W'$ also include a $Z'$ boson. 
If that $Z'$ couples to the SM leptons, as in left-right symmetric models, then the dilepton resonance searches 
force the $Z'$ to be significantly heavier than the $W'$. This constrains the extended Higgs sector responsible for their masses.

\medskip

{\it \bf $\boldsymbol{W'}$ interactions with quarks.}---A  $W'$ boson produced in the $s$ channel with a large cross section must couple to 
first generation quarks.  In order to avoid large flavor-changing neutral currents, it is natural to assume that the couplings are 
approximately flavor diagonal:
\be
\frac{g_{_{\rm R}} }{\sqrt{2} } W'^{+}_\mu \left( \bar u_R \gamma^\mu d_R + \bar c_R \gamma^\mu s_R + \bar t_R \gamma^\mu b_R   \right)  + {\rm H.c.}
\label{eq:quarks}
\ee
The $g_{_{\rm R}} \! $ parameter   can be extracted from cross section measurements for the dominant decay modes.
The widths for the $W'$ decays into $jj$ and $t\bar b$ are given by
\be
\Gamma (W'\to jj) \simeq 2 \Gamma(W'\to t \bar b) \simeq  \frac{g_{_{\rm R}}^2}{8\pi} M_{W'} ~~.
\label{eq:width-jets}
\ee
The $W'$ production cross section $\sigma (W')$ is $(g_{_{\rm R}}/g)^2$ times the SM rate for a heavier $W$,
where  $g \simeq 0.65$ (at 2 TeV) is the SM $SU(2)_W$ gauge coupling. 
Fig.~1 shows the next-to-leading order (NLO) cross sections at the LHC for $g_{_{\rm R}} \!= 0.5$.
We obtained these 
 by multiplying the leading-order cross sections computed with MadGraph  \cite{Alwall:2014hca} (using model files generated with 
FeynRules \cite{Alloul:2013bka} and CTEQ6L parton distributions \cite{Pumplin:2002vw} with factorization and renormalization scales set at $M_{W'}$) 
by scale-dependent $K$-factors. These are computed in \cite{Cao:2012ng}, and are in the  1.32--1.37 range for $\sqrt{s} = 8$ TeV 
(1.25--1.28 range for $\sqrt{s} = 13$ TeV) when  $M_{W'}$ varies from  $1.7$ to $2.2$ TeV. 
At 8 TeV, $\sigma(W') \approx 350$ fb (175 fb) for $M_{W'} = 1.8$ TeV (2 TeV).

\begin{figure}[t] 
\begin{center} 
\caption{NLO cross sections for $W'$ production at $\sqrt{s}=8$,  13 and 14 TeV,   for  $g_{_{\rm R}} = 0.5$. The cross sections scale as $g_{_{\rm R}}^2$.
 \\ [-1.1mm] } 
 \hspace*{-0.4cm} \includegraphics[width=0.44\textwidth, angle=0]{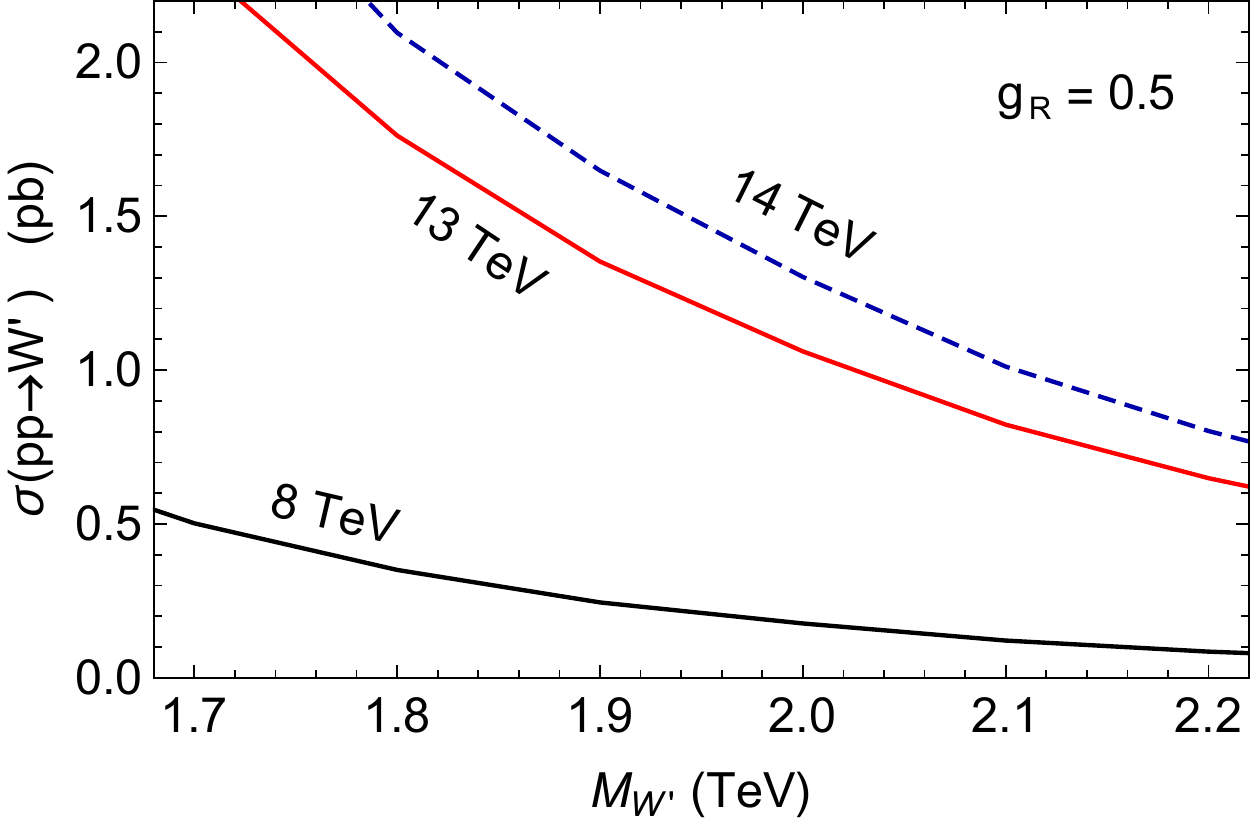}
\label{fig:sigma}
 \vspace*{-0.6cm}
\end{center}
\end{figure}

The CMS dijet excess requires a cross section times geometric acceptance ($A_{jj}$) roughly in the 50--100 fb range (the 95\% CL upper limit is
150 fb \cite{Khachatryan:2015sja,Aad:2014aqa}). Our simulation gives $A_{jj} \approx  47\%$, so that 
\be
\sigma_{jj}(W') \equiv  \sigma(pp \to W' \to jj)  \sim 100\!\!-\!200 \, {\rm fb} ~~.
\label{eq:dijet}
\ee
It follows that $g_{_{\rm R}} \! \gtrsim 0.4$
for $M_{W'} \! = 1.8$ TeV ($g_{_{\rm R}} \gtrsim 0.5$  for $M_{W'} = 2$ TeV); this lower limit 
corresponds to the case where the  $jj$ and $t\bar b$ channels saturate the total width. 

The fitted dijet rate implies that the predicted rate for $pp \to W' \to  t b $ is  $\sigma_{jj}/2 \sim 50 \! - \! 100$ fb at $\sqrt{s} = 8$ TeV, 
which is 
below  the sensitivity achieved by ATLAS searches in this channel \cite{Aad:2014xra}, but in tension with the CMS limits \cite{Chatrchyan:2014koa}.
This rate increases by a factor of 5 at $\sqrt{s} = 13$ TeV, allowing a definitive test for the presence of a  $tb$ peak near 2 TeV.

\medskip

{\it \bf $\boldsymbol{ W' \to WZ}$ signals.}---The  $W'$ coupling to $WZ$ arises from the kinetic terms of an extended 
gauge sector, such as $SU(2)\times SU(2)\times U(1)$, and takes the form
\bear
\frac{g_{_{\rm R}}}{c_W}  \,  \xi_Z    \frac{M_W^2}{M_{W'}^2} \,  i \, &&
 \left[ W^{\prime + }_\mu \left(W^-_\nu \partial^{[\nu}Z^{\mu]} +  Z_\nu \partial^{[\mu} W^{-\nu]}  \rule{0pt}{12pt}  \right)
    \right.
\nonumber \\ 
&&  \left. 
\; \;\; + \, Z_\nu W^-_\mu \partial^{[\nu} W^{\prime +\mu]}   \rule{0pt}{14pt} \right]  +  {\rm H.c.} ~~,
\label{eq:WZ}
\eear
where $c_W \equiv \cos\theta_W \approx 0.88$, and 
$[\mu,~\nu]$ represents commutation of indices ($\mu\nu-\nu\mu$).  
The factor of $(M_W/M_{W'})^2$ is due to $W-W'$ mass mixing, and the $\xi_Z$ coefficient is of order one. 
The $W' \to WZ$ width is given by 
\be 
\Gamma (W'\to WZ) = \frac{g_{_{\rm R}}^2 \, \xi_Z^2 }{192\pi } M_{W'}   ~~.
\label{eq:width-WZ}
\ee

The $pp \to W' \to WZ$ cross section, $\sigma_{WZ}(W')$,  is predicted in terms of the $jj$ one based on Eqs.~(\ref{eq:width-jets}) and (\ref{eq:width-WZ}): 
\be
\frac{\sigma_{WZ}(W')}{ \sigma_{jj}(W')} = \frac{\Gamma (W'\to WZ)}{\Gamma (W'\to jj)}  = \frac{ \xi_Z^2}{24}  ~~.   
\ee
Using Eq.~(\ref{eq:dijet}), we find $\sigma_{WZ}(W') \approx (4\!\!-\! 8)\,$fb$\,\times\, \xi_Z^2\,$.

The ATLAS search for $pp \! \to W' \! \to WZ \! \to JJ$ has identified 13 events with $JJ$ mass in the 1.85--2.05 TeV range, where the background is
5 events (Fig.~5a of \cite{Aad:2015owa}). The event selection efficiency is between 0.10 and 0.16 (Fig.~2b of \cite{Aad:2015owa}), implying
$\sigma_{WZ}(W') \approx  3\!-\! 10$ fb. Comparing this measured range with the predicted  $\sigma_{WZ}(W')$ we find
$0.6 \lesssim \xi_Z \lesssim 1.6$. Values of $\xi_Z$ in the 0.6--1 range are natural in simple Higgs sectors,  
and are allowed by the electroweak observables due to the $(M_W/M_{W'})^2$ suppression \cite{Dobrescu:2015yba}. 
Other explanations for the $JJ$ peak are discussed in \cite{Fukano:2015hga}.

It is imperative to check that  the ATLAS $WZ \to JJ$ peak is consistent with results obtained in other $WZ$ final states searched at the LHC.
Semileptonic final states of $W' \to WZ$ are particularly sensitive. The case where $W \to \ell \nu $ and $Z \to q\bar q$ 
is constrained by a CMS search \cite{Khachatryan:2014gha} optimized for a bulk graviton that decays to $WW$. 
At first sight there appears to be some conflict \cite{Pierini} with the ATLAS $WZ \to JJ$ signal. 
However,  a $1\sigma$ upward fluctuation in the cross section limit (Fig.~9 of \cite{Khachatryan:2014gha})  for a mass of 1.8 TeV
relaxes that conflict. In addition, the upper limit of 6 fb on the cross section for bulk graviton
production translates into an upper limit on $W'$ production that is higher by a factor of 2.2;
this is due to the lack of a combinatorial factor of 2 in the $WZ$ final state compared to the $WW$ one,
and also due to the $b$ veto imposed on the $WW$ search. As a result,  $\sigma_{WZ}(W') < 13$ fb at the 95\% CL. 
The ATLAS search \cite{Aad:2015ufa}  for  $W' \to WZ \to \ell\nu J$ also imposes $\sigma_{WZ}(W') < 13$ fb.
Thus, values of $\sigma_{WZ}(W')$ in the 3--10 fb range remain viable. 

The case where $Z \to \ell^+ \ell^- $ and $W$ decays to quarks
is constrained by the CMS search \cite{Khachatryan:2014gha} 
for a bulk graviton that decays to $ZZ$. 
The expected limit on the rate shown in Fig.~9 of \cite{Khachatryan:2014gha} is 7 fb for a mass in the 1.8--1.9 TeV range. Interestingly, the observed 
limit is $2\sigma$ weaker (around 15 fb), adding one more channel to the list of excesses near 2 TeV. 
The  $W'\to WZ$ semileptonic signal that would account for this $\sim 2\sigma$ excess 
is compatible with the $JJ$  excess (notice a combinatorial factor of 2). 

\medskip

{\it \bf $\boldsymbol{ W' \to Wh^0}$ signals.}---The kinetic terms of the extended Higgs sector responsible for breaking the 
$SU(2)\times SU(2)\times U(1)$ gauge symmetry include
a $W^\prime W h^0$ interaction term given by
\be
- g_{_{\rm R}} \, \xi_h  \, M_W \, W^{\prime\pm}_\mu W^{\mu \mp} h^0 ~~ ,
\label{eq:Wh}
\ee
where $\xi_h$ is a parameter of order one that depends on the details of the Higgs sector. The width for $W^\prime \to Wh^0 $ is 
\be
\Gamma (W'\to W h^0) = \frac{g_{_{\rm R}}^2 \, \xi_h^2 }{192\pi}  M_{W'}  ~~.
\label{eq:width-Wh}
\ee
If the SM Higgs doublet does not mix  with other fields, then $\xi_h = \xi_Z$ and 
$\Gamma (W'\! \to W h^0) \simeq \Gamma (W'\!\to W Z)$, as required by the equivalence theorem. 
The agreement between the SM and the measured $h^0$ properties indicates that the deviations from $\xi_h = \xi_Z$ are small.

In this case the $pp \to W' \!\to Wh^0$ cross section satisfies $\sigma_{Wh}(W') \approx \sigma_{WZ}(W')$.
Searches for $W' \!\to Wh^0 \!\to \ell\nu b\bar b$ should yield a signal comparable to that for $W' \!\to WZ \!\to JJ$ times 
$B(Wh^0\!\to \ell\nu b\bar b) $$/ B(WZ\!\to 4j) \approx 0.27$. The 8 excess $JJ$  events reported by ATLAS imply that there should be  a few 
excess $\ell\nu b\bar b$ events
(the $\ell\nu b\bar b$ selection efficiency depends on the efficiency for $h^0$ tagging, which we estimate  to be similar to the one for 
$WZ$ tagging). 
The CMS  $W' \!\to\! Wh^0$ search has reported 3 $\ell\nu b\bar b$  events in the 1.8--1.9 TeV mass bin for a background of 0.3.
This supports the assumption that the $\ell\nu b\bar b$ and $JJ$ excess events originate from a $W'$ boson.

The small number of events observed in these channels implies large  uncertainties. These can be reduced by 
searches in similar channels. We note here only the CMS search \cite{Khachatryan:2015bma} 
for $ W' \!\to Wh^0$ in hadronic final states ($6j$ and $bbjj$), which exhibits a small ($1\sigma$) excess at $M_{W'} \approx 1.8$ TeV,
setting a $\sigma_{Wh}(W') < 18$ fb limit  at 95\% CL.

\bigskip

{\it \bf Leptonic $\boldsymbol{W'}$ decays.}---The $W'$ considered here does not directly couple to left-handed leptons, implying highly suppressed
$W'$ decays into SM $\ell\nu$ pairs  (due to the small $W-W'$ mixing).
In order to fit the CMS $eejj$ excess, and to avoid large flavor-changing effects, we assume $W'$ coupling to leptons approximately given by
\be
\frac{g_{_{\rm R}} }{\sqrt{2} } \, W'^+_\nu  \! \left( \overline N_R^e \gamma^\nu  e_R + \overline N_R^\mu  \gamma^\nu \mu_R    
+ \overline N_R^\tau \gamma^\nu  \tau_R  \right)   + {\rm H.c.} ~,
\label{eq:leptons}
\ee
with the heavy right-handed neutrinos ($N^e_R,N^\mu_R,N^\tau_R$) being part of three vectorlike fermions with Dirac masses. 
Since the CMS $\mu\mu jj$ search \cite{Khachatryan:2014dka}
has not yielded deviations from the SM, the $N^\mu$ mass must satisfy $m_{N^\mu} > M_{W'}$. 

The $N^\tau$ fermion can be light because 
no dedicated $W' \!\to \tau N^\tau \!\to \tau\tau jj$ search has been performed.
$N^\tau$ may even couple to the electron or muon \cite{Aguilar-Saavedra:2014ola}:
\be
\frac{g_{_{\rm R}} }{\sqrt{2} }  W'^+_\nu \, \overline N_R^\tau  \gamma^\nu  \left(  s_{\theta_e}   e_R   +  s_{\theta_\mu}  \mu_R \right)   + {\rm H.c.}
\ee
with $s_{\theta_\mu} \! < s_{\theta_e} \lesssim 0.5$ 
leads to  $W' \! \to e\tau jj$ or $\mu\tau jj$ signals that have escaped detection,  
and slightly decreases  the diagonal couplings (\ref{eq:leptons}).
In that case  an $e^+e^-jj$ signal is produced by $W' \!\to e N^\tau$, so  $N^e$ may also be heavier than $W'$.
The $W'^+$ decay into $e^+N^\tau$ has a width
\be \hspace*{-2.4mm}
\Gamma (W' \!\!\to e N^\tau) = \frac{g_{_{\rm R}}^2  s_{\theta_e}^2 \! }{48\pi} M_{W'} \! \left( \! 1 \! + \!\frac{m_{N^\tau}^2\!}{2 M_{W'}^2 \! } \! \right)
\! \left( \! 1\!  - \! \frac{m_{N^\tau}^2\!}{M_{W'}^2 }  \!\right)^{\! 2}  \! \!. \!\!
\label{eq:width-leptons}
\ee

The  $B(N^\tau \to ejj)$ branching fraction is naively about $0.6 \, s_{\theta_e}^2$. 
However, $N^\tau$ decays into $e t \bar b$ with 
hadronic top decays, or into $e W Z/h^0$ with hadronic decays of SM bosons also appear as $ejj$, especially for
boosted topologies; effectively,  $B(N^\tau\! \to e jj) \sim 0.9 s_{\theta_e}^2$. 
The  $pp\! \to    W' \!\to  eN^\tau\!   $$ \to e^+e^-jj$ rate, $\sigma_{eejj}(W')$, is smaller than the $jj$ signal by a  $B(N^\tau \!\to ejj) \Gamma (W' \!\to eN^\tau)/\Gamma (W' \!\to jj)$ factor.

The $eejj$ excess requires $\sigma_{eejj}(W')$  roughly in the 1--2 fb range (see Fig.~4 of \cite{Khachatryan:2014dka}), 
so that it is 0.5--2\% of the dijet signal.
For $m_{N^\tau} \sim 1$ TeV and $s_{\theta_e} \approx 0.5$, we find a predicted ratio 
$\sigma_{eejj}(W')/\sigma_{jj}(W') \approx 0.6\%$,
consistent with the signal rates indicated by the data. 

The $e\tau jj$ final state produced by  $W' \to eN^\tau, \tau N^\tau$ is also interesting. 
The hadronic $\tau$ decay leads to an $e+\met$ + jets signal that may explain the $2.6\sigma$
CMS excess reported in \cite{CMS:2014qpa}. 
The leptonic $\tau$ decays modify the ``flavor-symmetric" background,
which distorts the  kinematics of the $eejj$ signal, potentially in agreement with observations made in \cite{Khachatryan:2014dka}. 
An alternative is $m_{N^\tau}< m_{N^e} < M_{W^\prime}$. The $N^e$-$N^\tau$ mixing then leads  to two $e^+e^-jj$ contributions, with
$ejj$ distributions peaked at different masses.

\bigskip

{\it \bf A baseline $\boldsymbol W'$ model.}---Let us summarize the $W'$ model introduced so far. 
The primary parameters are $M_{W'}$, $g_{_{\rm R}}$, $\xi_Z \approx \xi_h$, $m_{N^\tau}$, $s_{\theta_e}$.
The masses of $N^e$ and $N^\mu$ are above $M_{W'}$ and are not relevant here;
the coupling of $W'$ to $\mu N^\tau$, $s_{\theta_\mu} $, is a  parameter 
that could become relevant if $W'$ processes with muons are observed.

The mass peaks for $jj$, $Wh^0$, and $WZ \to J \ell\ell$ indicate $M_{W'} \approx 1.8$--1.9 TeV, while the 
$WZ \to JJ$ peak is around 1.9--2.0 TeV. The relatively low resolution and
the small number of  events  makes it likely that the $JJ$ peak would migrate towards 1.85 TeV with more data,
if a $W'$ boson exists.
The cross sections consistent with the $WZ \to JJ$ and $Wh^0$ peaks require $\xi_Z \approx \xi_h \approx $ 0.6--1 for simple Higgs sectors.
The $W'eN^\tau$ coupling is  $s_{\theta_e} \approx$ 0.4--0.5
in order to explain the $eejj$ signal.
The $N^\tau$ mass is loosely constrained, $m_{N^\tau} \sim $ 0.4--1.2
 TeV. 

Some $W'$ decays could 
involve scalars from the extended Higgs sector \cite{Dobrescu:2013gza}, or other new particles. Let $B_X$ be their combined branching fraction.
For $B_X = 0$, the $W'$ branching fractions are $B( jj) = 2 B( t\bar b) \approx 60\%$, $B(WZ) \approx B( Wh^0) \approx 2\%$,
$B( eN_\tau)\approx 1.5\%$,  $B( \tau N_\tau)\approx 4.5\%$. 
The cross section that can account for the $jj$ peak then implies $g_{_{\rm R}} \approx 0.45 - 0.6$.
For $B_X >0 $, $g_{_{\rm R}}$ scales as $(1- B_X)^{-1/2}$, so that the left-right symmetric relation $g_{_{\rm R}} = g$ is recovered for 
$B_X \sim 20\%$--50\%.

\bigskip

{\it \bf An {\small  $\boldsymbol{SU(2)_L\times SU(2)_R \times U(1)_{B-L}}$} theory.}---We now present a renormalizable theory that embeds
our baseline model. Any gauge symmetry associated with a $W'$ also involves a $Z'$ with correlated properties. The limits on dilepton resonances require a Higgs sector that 
allows $M_{Z'} \gtrsim 1.5 M_{W'}$.
 In the original $SU(2)_L\times SU(2)_R \times U(1)_{B-L}$ theory \cite{Pati:1974yy,Mohapatra:1986uf} the right-handed neutrinos may be very heavy only if they 
 have Majorana masses, which 
(barring tiny mass splittings \cite{Gluza:2015goa}) leads to same-sign $\ell^\pm\ell^\pm jj$ events, in contradiction to the CMS result.

In order for $N^\tau_R$ to acquire a Dirac mass we introduce a vectorlike fermion $\psi = (\psi^N, \psi^\tau)^\top $ transforming as $(2,+1)$ under  $SU(2)_R \times U(1)_{B-L}$.
Its $\psi_L^N$ component can become the Dirac partner of $N_R^\tau$. 
To see that, let us first describe a simple Higgs sector: an  $SU(2)_R$ triplet scalar $T$ breaks $SU(2)_R\times U(1)_{B-L} $ to $U(1)_Y$ 
giving the bulk of $M_{W'}$ and $M_{Z'}$,  and a 
bidoublet scalar $\Sigma$ breaks $SU(2)_L\times U(1)_Y \to U(1)_Q$ inducing a small mixing between the charged gauge bosons.
For $M_{W'}\gg M_W$, $\Sigma$ consists of two $SU(2)_W$ Higgs doublets, which break the electroweak symmetry.
The SM Higgs does not mix with other scalars in the alignment limit, and the other charged and neutral scalars could be at the TeV scale. 

A large Majorana mass for $\psi_R^N$ arises from the $\bar \psi_R^c T^\dagger \psi_R$ coupling. 
Below the $\psi_R^N$ mass, a Dirac mass for $N_R^\tau$ and $\psi_L^N$  is generated
by the $\bar \psi_L T (N_R^\tau,\tau_R)^\top$ coupling. Finally, $\psi^\tau$  gets a mass from a $\bar \psi_L\psi_R$ term. The latter also induces
a  contribution to the mass  of $\psi^N$, which cannot be much larger than $m_{N^\tau}$. 
Thus, the charged fermion $\psi^\tau$ is expected to have an $O(M_{W'})$ mass. The same mechanism may involve
Dirac partners for $N^e_R$ and $N^\mu_R$. 

With the field content of this theory shown in Table~\ref{table:SU}, the fermion kinetic terms induce the $W^\prime$ couplings to quarks and leptons discussed earlier, and
$g_{_{\rm R}} \!$ from Eq.~(\ref{eq:quarks}) 
is the $SU(2)_R$ gauge coupling up to corrections of order $M_W^2/M_{W'}^2$.
Comparing the bosonic kinetic terms with the $W^\prime WZ$ and $W^\prime Wh^0$ couplings of Eqs.~(\ref{eq:WZ}) and (\ref{eq:Wh}), 
we find $\xi_h=\xi_Z=\sin2\beta$  in the Higgs alignment limit \cite{Dobrescu:2015yba}, where $\tan\beta$ is the ratio of the two $\Sigma$ VEVs.

\medskip

\begin{table}[t]
\renewcommand{\arraystretch}{1.25}
\caption[]{\ $SU(2)_L \times SU(2)_R \times U(1)_{B-L}$ gauge charges. The SM fermions have generation-independent charges.}  
\vspace*{3mm}
\centering
\begin{tabular}{|c|c|c|c|}
\hline
Fields & $SU(2)_L$ & $SU(2)_R$ & $U(1)_{B-L}$ \\ \hline\hline
$(u_L,~d_L)$ & 2 & 1 & +1/3 \\ 
$(u_R,~d_R)$ & 1 & 2 & $+1/3$ \\ 
$(\nu_L,~\ell_L)$ & 2 & 1 & $-1$ \\ 
$(N_R,~\ell_R)$ & 1 & 2 & $-1$ \\ 
$\psi_L$ , $\psi_R$ & 1 & 2 & $+1$ \\  \hline 
$\Sigma$ & 2 & 2 & 0 \\ 
$T$ & 1 & 3 & +2 \\ \hline
\end{tabular}
\vspace*{-1mm}
\label{table:SU}
\end{table}

{\it \bf Predictions for the $Z'$ boson.}---The $Z'$ boson  is an $SU(2)_R\times U(1)_{B-L}$ gauge boson, with a small $SU(2)_L$
admixture governed by $M_Z^2/M_{Z^\prime}^2$. The $Z^\prime$ mass is 
\begin{equation}
M_{Z^\prime} =  \sqrt {2} \, g_{_{\rm R}} \left(g_{_{\rm R}}^2 -g^{\prime 2}\right)^{-1/2} M_{W^\prime}  ~~,
\end{equation}
where $g' \approx 0.36$ is the hypercharge  gauge coupling.
This implies $M_{Z^\prime} >1.5 M_{W^\prime}$ as a consequence of the large  $SU(2)_R$-breaking VEV of the $T$ scalar. 
The value of $g_{_{\rm R}}$ indicated by the excess events attributed to $W'$ further constrains $M_{Z^\prime}/M_{W^\prime}$. For 
$M_{W^\prime} =1.9$ TeV, the preferred range of $0.45 < g_R < 0.6$ implies 3.4 TeV $< M_{Z'} <$ 4.5 TeV. A 
larger $g_{_{\rm R}}$ due to $B_X > 0$ 
would slightly reduce the lower limit on $M_{Z'}$.

\begin{figure}[t] 
\begin{center}
\caption{$Z'$ production cross section times branching fractions as a function of $M_{Z'}$, for $M_{W^\prime} \!=1.9$ TeV at the 13 TeV LHC. Shaded bands correspond
to $M_{W^\prime}$ in the 1.8--2.0 TeV range.  \\ [-2mm] } 
 \hspace*{-0.4cm}
\includegraphics[width=0.45\textwidth, angle=0]{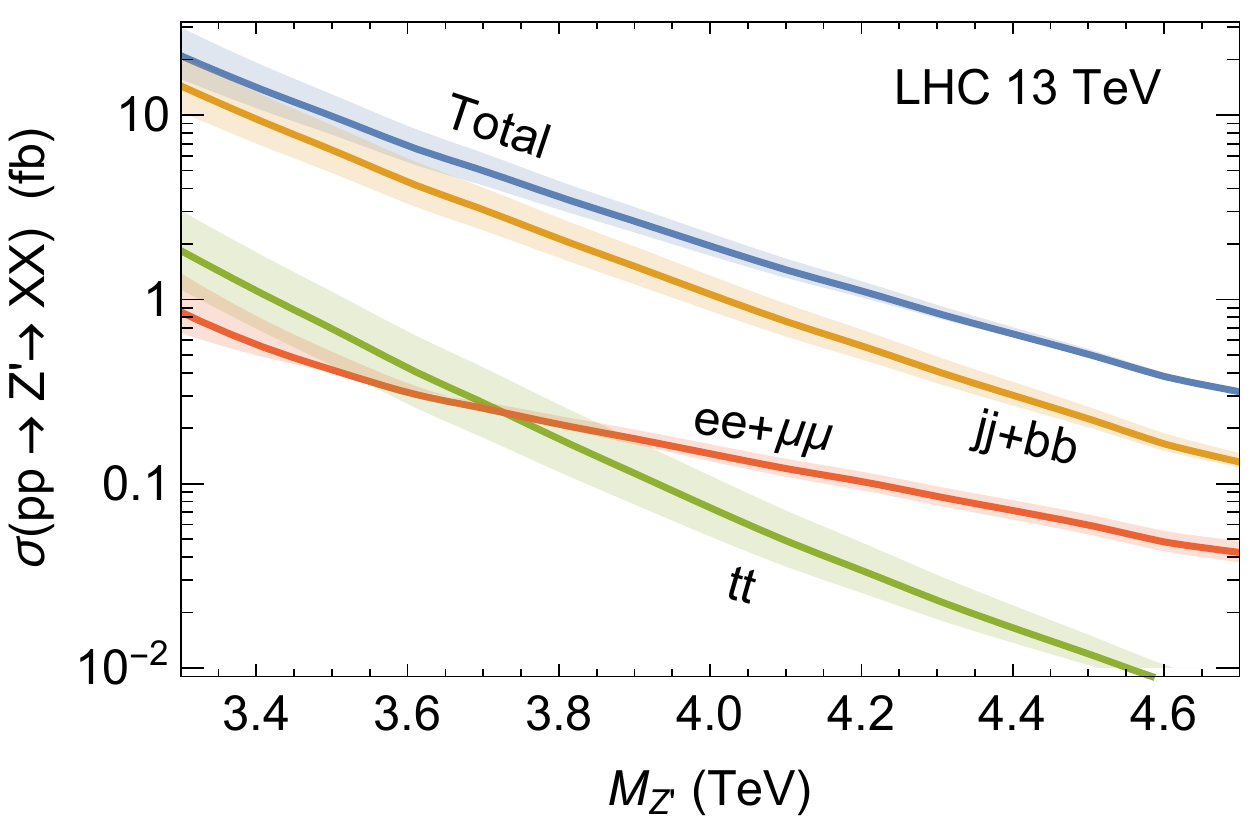}
 \vspace*{-0.6cm}
\label{fig:zprime}
\end{center}
\end{figure}

The fermion couplings to $Z^\prime$ are given by 
\be
 \Big(g_{_{\rm R}}^2 T^3_R -  g_{B-L}^2 \frac{B\!-\!L}{2} \Big) \left(g_{_{\rm R}}^2+ g_{B-L}^2\right)^{-1/2}  ~~.
\ee
The $U(1)_{B-L}$ gauge coupling is also determined by $g_{_{\rm R}}$: \
 $g_{\tiny B-L}  = (1/g^{\prime 2}-1/g_{_{\rm R}}^2)^{-1/2}$.
Thus, the theory is highly predictive, {\it e.g.}, $M_{W^\prime}$ and $M_{Z^\prime}$ measurements would fix the $Z'$ couplings.
Fig.~\ref{fig:zprime} shows $Z^\prime$ production cross section 
times branching fractions at the 13 TeV LHC  for 
$M_{W^\prime} \! = 1.8$--2 TeV, $m_{N^\tau} \!= 1$ TeV and 
$m_{N^e}, m_{N^\mu} \! > \! M_{Z'} \! /2$.
The $Z^\prime$ production rate computed using MadGraph 5 is multiplied in Fig.~\ref{fig:zprime} by a constant $K$ factor of 1.2.  
Besides the decay modes shown there (dijet, $\ell^+\ell^-, t\bar t$), several others are phenomenologically important, including  $W^+W^-$, $Z h^0$, 
$N^\tau \bar{N^\tau}$.

\medskip

{\it \bf Conclusions.}---The $W'$ model presented here appears to be a viable description of the small mass peaks near 2 TeV
observed in at least five channels at the LHC. Definitive tests of this model will be performed in several $W'$ decay channels in Run 2 of the LHC.
Assuming an $SU(2)_L\times SU(2)_R\times U(1)_{B-L}$ gauge origin of the $W'$, we  predict the existence of a $Z'$ boson of mass below 4.5 TeV with 
production rates shown in Fig.~2. Our renormalizable theory includes Dirac masses for  right-handed neutrinos.

\bigskip

{\it \bf Acknowledgments:} We thank 
Patrick Fox, Robert Harris, Ian Lewis, Maurizio Pierini and Nhan Tran 
for constructive comments. ZL was supported by the Fermilab Graduate Student Research Program in Theoretical Physics.

 \vfil 

\begin{thebibliography}{99} 

\bibitem{Aad:2015owa} 
  G.~Aad {\it et al.}  [ATLAS Collaboration],
  ``Search for high-mass diboson resonances with boson-tagged jets in $pp$ collisions at $\sqrt{s}$ = 8 TeV," 
  arXiv:1506.00962.

\bibitem{Khachatryan:2014hpa} 
  V.~Khachatryan {\it et al.}  [CMS Collaboration],
``Search for massive resonances in dijet systems containing jets tagged as W or Z boson decays in pp collisions at $ \sqrt{s} $ = 8 TeV,''
  JHEP {\bf 1408}, 173 (2014)
 [arXiv:1405.1994].

\bibitem{Khachatryan:2014dka} 
  V.~Khachatryan {\it et al.}  [CMS Collaboration],
  ``Search for heavy neutrinos and $\mathrm {W}$ bosons with right-handed couplings in $pp$ collisions  at $\sqrt{s} = 8\,\text {TeV} $,''
  Eur.\ Phys.\ J.\ C {\bf 74}, no. 11, 3149 (2014)
  [arXiv:1407.3683 [hep-ex]].

\bibitem{CMS:2015gla} 
  CMS Collaboration,
  ``Search for massive $WH$ resonances decaying to $\ell \nu b \bar{b}$ final state in the boosted regime at $\sqrt{s}=8$ TeV,''
  note PAS-EXO-14-010, March 2015.

\bibitem{Khachatryan:2015sja} 
  V.~Khachatryan {\it et al.}  [CMS Collaboration],
  ``Search for resonances and quantum black holes using dijet mass spectra in $pp$ collisions at $\sqrt{s} =$ 8 TeV,''
  Phys.\ Rev.\ D {\bf 91}, no. 5, 052009 (2015)
  [arXiv:1501.04198 [hep-ex]].

\bibitem{Aad:2014aqa} 
  G.~Aad {\it et al.}  [ATLAS Collaboration],
  ``Search for new phenomena in the dijet mass distribution using $pp$ collision data at $\sqrt{s}=8$ TeV,'' 
  Phys.\ Rev.\ D {\bf 91}, no. 5, 052007 (2015)
  [arXiv:1407.1376 [hep-ex]]; 

\bibitem{Mohapatra:1986uf} 
See {\it e.g}, R.~N.~Mohapatra,
  ``Unification and supersymmetry. The frontiers of quark - lepton physics,''
  New York, USA: Springer (2003) 421 p; Ch. 6.

\bibitem{Keung:1983uu} 
  W.~Y.~Keung and G.~Senjanovic,
  ``Majorana neutrinos and the production of the right-handed charged gauge boson,''
  Phys.\ Rev.\ Lett.\  {\bf 50}, 1427 (1983).
  
\bibitem{Gluza:2015goa} 
  J.~Gluza and T.~Jelinski,
  ``Heavy neutrinos and the $pp\to lljj$ CMS data,''
  arXiv:1504.05568.
  
\bibitem{Alwall:2014hca} 
  J.~Alwall {\it et al.},  
  ``The automated computation of tree-level and NLO differential cross sections, and their matching to parton shower simulations,''
  JHEP {\bf 1407}, 079 (2014)
  [arXiv:1405.0301].

\bibitem{Alloul:2013bka} 
  A.~Alloul, N.~D.~Christensen, C.~Degrande, C.~Duhr and B.~Fuks,
  ``FeynRules  2.0 - A complete toolbox for tree-level phenomenology,''
  Comput.\ Phys.\ Comm.\  {\bf 185}, 2250 (2014)
  [arXiv:1310.1921].
  
\bibitem{Pumplin:2002vw} 
  J.~Pumplin {\it et al.},   
  ``New generation of parton distributions with uncertainties from global QCD analysis,''
  JHEP {\bf 0207}, 012 (2002)
  [hep-ph/0201195].
    
\bibitem{Cao:2012ng} 
  Q.~H.~Cao, Z.~Li, J.~H.~Yu and C.~P.~Yuan,
  ``Discovery and identification of $W'$ and $Z'$ in $SU(2) \times SU(2) \times U(1)$ models at the LHC,''
  Phys.\ Rev.\ D {\bf 86}, 095010 (2012)
  [arXiv:1205.3769].

\bibitem{Aad:2014xra} 
  G.~Aad {\it et al}  [ATLAS Collaboration],
  ``Search for $W' \!\!\to tb \!\rightarrow \! qqbb$ decays,"  
  Eur.\ Phys.\ J.\ C {\bf 75}, no. 4, 165 (2015)
  [arXiv:1408.0886];
  ``Search for $W' \to t\bar{b}$ in the lepton plus jets final state," 
  Phys.\ Lett.\ B {\bf 743}, 235 (2015)
  [arXiv:1410.4103]. 

\bibitem{Chatrchyan:2014koa} 
  S.~Chatrchyan {\it et al.} [CMS Collaboration],
  ``Search for $W' \!\to tb$ decays in the lepton + jets final state in pp collisions at $\sqrt{s}$ = 8 TeV,''
  JHEP {\bf 1405}, 108 (2014)
  [arXiv:1402.2176];
  ``Search for W' to tb in $pp$ collisions at $\sqrt{s} = 8$ TeV,''
  arXiv:1509.06051.


\bibitem{Dobrescu:2015yba} 
  B.~A.~Dobrescu and Z.~Liu,
  ``Heavy Higgs bosons and the 2 TeV $W'$ boson,''
  arXiv:1507.01923.

\bibitem{Fukano:2015hga} 
  H.~S.~Fukano {\it et al.}, 
  ``2 TeV Walking Technirho at LHC?,''
  arXiv:1506.03751. 
  J.~Hisano, N.~Nagata and Y.~Omura,
  ``Interpretations of the ATLAS Diboson Resonances,''
  arXiv:1506.03931. 
  D.~B.~Franzosi, M.~T.~Frandsen and F.~Sannino,
  ``Diboson signals via Fermi scale spin-one states,''
  arXiv:1506.04392. 
  K.~Cheung, W.~Y.~Keung, P.~Y.~Tseng and T.~C.~Yuan,
  ``Interpretations of the ATLAS diboson anomaly,''
  arXiv:1506.06064.

\bibitem{Khachatryan:2014gha} 
  V.~Khachatryan {\it et al.}  [CMS Collaboration],
 ``Search for massive resonances decaying into pairs of boosted bosons in semi-leptonic final states at $\sqrt{s} =$ 8 TeV,''
  JHEP {\bf 1408}, 174 (2014).
  [arXiv:1405.3447].

\bibitem{Pierini}
M. Pierini, talk at Fermilab Users Meeting, June  2015.

\bibitem{Aad:2015ufa} 
  G.~Aad {\it et al.}  [ATLAS Collaboration],
  ``Search for production of $WW/WZ$ resonances decaying to a lepton, neutrino and jets in $pp$ collisions at $\sqrt{s}=8$  TeV,''
  Eur.\ Phys.\ J.\ C {\bf 75}, no. 5, 209 (2015)
  [arXiv:1503.04677].

\bibitem{Khachatryan:2015bma} 
  V.~Khachatryan {\it et al.}  [CMS Collaboration],
  ``Search for a massive resonance decaying into a Higgs boson and a $W$ or $Z$ boson in hadronic final states in $pp$ collisions at $\sqrt{s} = 8$ TeV,''
  arXiv: 1506.01443.

\bibitem{Aguilar-Saavedra:2014ola} 
  J.~A.~Aguilar-Saavedra and F.~R.~Joaquim,
  ``Closer look at the possible CMS signal of a new gauge boson,''
  Phys.\ Rev.\ D {\bf 90}, no. 11, 115010 (2014)
  [arXiv:1408.2456].

\bibitem{CMS:2014qpa} 
  CMS Collaboration,
  ``Search for pair-production of first generation scalar leptoquarks in $pp$ collisions,''
note  PAS-EXO-12-041, July 2014.
 

\bibitem{Dobrescu:2013gza} 
  B.~A.~Dobrescu and A.~D.~Peterson,
  ``$W'$ signatures with odd Higgs particles,''
  JHEP {\bf 1408}, 078 (2014)
  [arXiv:1312.1999].
  A.~Jinaru  {\it et al.},    
  ``$W' \to hH^{\pm}$ decay in $G(221)$ models,''
  J.\ Phys.\ G {\bf 41}, 075001 (2014)
  [arXiv:1312.4268].

  \bibitem{Pati:1974yy}  J.~C.~Pati and A.~Salam,
  ``Lepton number as the fourth color,''
  Phys.\ Rev.\ D {\bf 10}, 275 (1974).
  R.~N.~Mohapatra and J.~C.~Pati,
  ``Left-right gauge symmetry and an isoconjugate model of CP violation,''
  Phys.\ Rev.\ D {\bf 11}, 566 (1975);
  ``A natural left-right symmetry,''
  Phys.\ Rev.\ D {\bf 11}, 
  2558 (1975).
  G.~Senjanovic and R.~N.~Mohapatra,
  ``Exact left-right symmetry and spontaneous violation of parity,''
  Phys.\ Rev.\ D {\bf 12}, 1502 (1975);
  R.~N.~Mohapatra and G.~Senjanovic,
  ``Neutrino mass and spontaneous parity violation,''
  Phys.\ Rev.\ Lett.\  {\bf 44}, 912 (1980).
  P.~Minkowski,
  ``$\mu \to e\gamma$ at a rate of one out of $10^{9}$ muon decays?,''
  Phys.\ Lett.\ B {\bf 67}, 421 (1977).

    
 \end{thebibliography}
\end{document}